\documentclass[sn-nature]{sn-jnl}

\usepackage{graphicx}%
\usepackage{multirow}%
\usepackage{amsmath,amssymb,amsfonts}%
\usepackage{amsthm}%
\usepackage{mathrsfs}%
\usepackage[title]{appendix}%
\usepackage{xcolor}%
\usepackage{textcomp}%
\usepackage{manyfoot}%
\usepackage{booktabs}%
\usepackage{algorithm}%
\usepackage{algorithmicx}%
\usepackage{algpseudocode}%
\usepackage{listings}%
\usepackage[T1]{fontenc}

\theoremstyle{thmstyleone}%
\theoremstyle{thmstyletwo}%

\theoremstyle{thmstylethree}%

\begin{document}

\title[Article Title]{A young gas giant and hidden substructures in a protoplanetary disk}


\author*[1]{\fnm{Álvaro} \sur{Ribas}}\email{ar2193@cam.ac.uk}

\author[2]{\fnm{Miguel} \sur{Vioque}}

\author[3]{\fnm{Francesco} \sur{Zagaria}}

\author[1]{\fnm{Cristiano} \sur{Longarini}}

\author[2]{\fnm{Enrique} \sur{Macías}}

\author[1]{\fnm{Cathie J.} \sur{Clarke}}

\author[4, 5, 6]{\fnm{Sebastián} \sur{Pérez}}

\author[7]{\fnm{John} \sur{Carpenter}}

\author[8]{\fnm{Nicolás} \sur{Cuello}}

\author[9]{\fnm{Itziar} \sur{de Gregorio-Monsalvo}}

\affil*[1]{\orgdiv{Institute of Astronomy}, \orgname{University of Cambridge}, \orgaddress{\street{Madingley Road}, \city{Cambridge}, \postcode{CB3 0HA}, \country{United Kingdom}}}

\affil[2]{\orgname{European Southern Observatory}, \orgaddress{\street{Karl-Schwarzschild-Str. 2}, \city{Garching bei München}, \postcode{85748}, \country{Germany}}}

\affil[3]{\orgname{Max Planck Institute for Astronomy}, \orgaddress{\street{Königstuhl 17}, \postcode{69117}, \city{Heidelberg}, \country{Germany}}}

\affil[4]{\orgdiv{Millennium Nucleus on Young Exoplanets and their Moons (YEMS)}, \orgaddress{\city{Santiago}, \country{Chile}}}

\affil[5]{\orgdiv{Departamento de Física, Universidad de Santiago de Chile}, \orgaddress{\street{Av. Victor Jara 3493}, \city{Santiago}, \country{Chile}}}

\affil[6]{\orgdiv{Center for Interdisciplinary Research in Astrophysics and Space
Science (CIRAS), Universidad de Santiago de Chile}, \orgaddress{\city{Santiago}, \country{Chile}}}

\affil[7]{\orgdiv{Joint ALMA Observatory}, \orgaddress{\street{Avenida Alonso de Córdova 3107, Vitacura}, \city{Santiago}, \country{Chile}}}

\affil[8]{\orgdiv{Univ. Grenoble Alpes, CNRS, IPAG}, \orgaddress{\city{Grenoble}, \postcode{38000}, \country{France}}}

\affil[9]{\orgdiv{European Southern Observatory}, \orgaddress{\street{Alonso de Córdova 3107, Casilla 19, Vitacura}, \city{Santiago}, \country{Chile}}}


\abstract{The detection of planets in protoplanetary disks has proven to be extremely challenging. In contrast, rings and gaps, usually attributed to planet-disk interactions, have been found in virtually every large protoplanetary (Class~II) disk observed at 0.9-1.3 mm with sufficient spatial resolution (5~au). The nearby disk around MP Mus (PDS~66) stands as an exception to this rule, and its advanced age (7-10~Myr) is particularly difficult to reconcile with its apparent lack of substructures. Despite the disk's smooth appearance, \textit{Gaia} data of MP~Mus show a significant proper motion anomaly, signalling the presence of a companion. Here we present ALMA 3~mm observations of the system with comparable high spatial resolution to previous 1.3~mm data. The new observations pierce deeper into the disk midplane and reveal an inner cavity ($<$3~au) and a ring at 10~au. The disk structure inferred from ALMA observations narrows down the properties of the companion to a gas giant orbiting at 1-3 au, and hydrodynamic simulations further confirm that such a planet can produce the observed cavity. These independent pieces of evidence constitute an indirect but compelling detection of an exoplanet within a protoplanetary disk using \textit{Gaia} astrometry. MP~Mus is the first system in which undetected substructures are revealed thanks to the lower optical depths at longer wavelengths, suggesting that rings and gaps are even more abundant than what is currently believed. 
}




\maketitle

\section{Main}\label{Main}

Despite extensive efforts to find protoplanets in protoplanetary disks, they have only been robustly detected in two systems so far \citep[PDS~70 and IRAS 04125+2902,][]{Keppler_2018, Haffert_2019,Barber_2024}. Accessing the population of young planets would be truly valuable for planet formation studies as this would provide direct benchmarks to our models, but common detection methods such as transits, radial velocity, or direct imaging are extremely challenging or simply unfeasible for embedded companions due to the high dust extinction in the disk and the increased stellar activity of young stars \citep[e.g.][]{Testi_2015,Donati_2020}. Instead, the presence of planets in disks is typically inferred by planet-disk interaction footprints such as disk substructures and velocity perturbations. In fact, arguably the most significant discovery of the last decade for the field of planet formation is the ubiquity of substructures in protoplanetary disks. Rings, gaps, spiral arms, or azimuthal asymmetries are found in virtually every large disk observed with sufficient angular resolution \citep{Andrews_2018,Long_2018,Bae_2023}.

Disk substructures offer a natural solution to the problem of dust radial drift: the interaction of mm/cm-sized grains with the gas in the disk causes large grains to drift towards the central star. This process is much faster than typical timescales for planet formation and can quickly deplete the population of large grains in disks, presenting a theoretical barrier to build rocky planets and the cores of gas giants \citep{Weidenschilling_1977,Brauer_2007}. Local gas pressure maxima (induced by planets or other processes) have been proposed as a mechanism to trap large grains and allow them to grow to larger sizes \citep{Paardekooper_2004,Lyra_2009,Pinilla_2012}, but this solution remained hypothetical until the high resolution and sensitivity of observatories such as the Atacama Large Millimeter/Submillimeter Array (ALMA) and the Spectro-Polarimetric High-contrast Exoplanet REsearch instrument (SPHERE) uncovered the plethora of disk substructures \citep[e.g.][]{Andrews_2018,Garufi_2024, Ginski_2024}.

As of today, MP~Mus is the only large Class~II protoplanetary disk that appears structureless when observed at millimetre wavelengths with high spatial resolution \citep[$<$5~au,][]{Ribas_2023}. This K1V star is located 97.9$\pm$0.1~pc away from Earth \citep{Gaia_2021b} and has an estimated age of 7-10~Myr \citep{Ribas_2023}. Its dust disk extends up to $R_{\rm dust, 90\%}$=45~au at 1.3~mm (i.e., the radius encompassing 90\% of the 1.3~mm continuum emission), while the gaseous disk traced by the $^{12}$CO(J=2-1) reaches $R_{\rm CO~90\%}$=110~au. This ratio of gas/dust radii is similar to that of structured disks, in which dust traps are likely acting \citep{Toci_2021}. However,  the disk appears structureless at 1.3~mm down to 4~au scales except for a tentative narrow dust ring in the outer regions \citep{Ribas_2023}, recently confirmed by \citep{Aguayo_2025} in high-resolution ALMA 0.89~mm observations. In contrast, scattered light observations of this system show an apparent gap between 40 and 80~au \citep{Wolff_2016, Avenhaus_2018}. Nevertheless, the fact that the millimetre disk extends up to $\sim$55~au, the clearly Keplerian rotation of the disk as traced by ALMA, and its flat appearance, strongly suggest that the drop in scattered light intensity between 40 and 80~au is due to a shadowing effect and not an actual gap in the disk \citep{Ribas_2023}. The smooth 1.3~mm radial profile is yet more puzzling when we compare the characteristic lifetime of disks \citep[$\sim$3~Myr,][]{Haisch_2001,Fedele_2010,Ribas_2015} with the age of MP~Mus. To explain the survival of large dust grains in such an old and seemingly structureless disk, \citep{Ribas_2023} proposed a number of possible solutions: 1) substructures may be present in the disk but the 1.3~mm continuum is optically thick, effectively hiding them, 2) substructures smaller than 4~au exist and remain unresolved with current observations, or 3) the disk is really featureless and a different (unknown) mechanism is retaining mm-emitting grains in it. None of these scenarios could be tested with the existing observations at the time, and MP~Mus remained an outlier among disks.

\subsection{The \textit{Gaia} proper motion anomaly of MP~Mus} \label{sec:PMa}

The 1.3~mm continuum emission of MP~Mus shows no obvious signs of planets in the system but, in contrast, its astrometry deviates from a single star solution. By measuring the proper motion vector of the photocenter at different times, we can identify changes (proper motion anomalies, PMas) that signal the presence of companions \citep{Perryman_2014, Kervella_2019, Kervella_2022}. PMas have already been used to identify exoplanets around main sequence stars \citep[later confirmed with high-resolution imaging, e.g.][]{deRosa2023, Mesa_2023, Currie_2023, Kiefer_2024}, but this method has not been applied to sources with protoplanetary disks yet. Compared to main sequence stars, protostars exhibit additional phenomena (e.g., variability, strong disk asymmetries, differential extinction, stellar obscuration by accretion columns) that could potentially induce a PMa. However, the analysis of well-known protoplanetary disks shows that only edge-on, highly asymmetric, or massive disks can induce proper motion anomalies (Vioque et al. in prep), and MP~Mus has a very low accretion rate \citep[$\sim10^{-10}$~$M_\odot$ yr$^{-1}$,][]{Ingleby_2023}. Therefore, none of these potential issues apply to the MP~Mus system \citep{Ribas_2023}. 

We determined the PMa of MP~Mus by comparing the proper motion vectors of \textit{Gaia}~DR2 and \textit{Gaia}~DR3 \citep[for details on the technique see][]{Penoyre_2022a, Penoyre_2022b}. The resulting PMa has a significance of 4.5~$\sigma$ and implies a change in the proper motion of $|\Delta \mu|=0.21\pm0.05$ mas~yr$~^{-1}$. We used the scanning law of \textit{Gaia}~DR2 [\citealp{Boubert_2021}] and \textit{Gaia}~DR3 \footnote{\href{https://gea.esac.esa.int/archive/documentation/GDR3/Gaia_archive/chap_datamodel/sec_dm_auxiliary_tables/ssec_dm_commanded_scan_law.html}{Gaia documentation}} to retrieve the epochs and scan angles in which \textit{Gaia} observed this source. We then used these epochs and angles, together with realistic \textit{Gaia} uncertainties, to simulate the \textit{Gaia}~DR2 and DR3 observations of MP~Mus assuming a two body system with different mass ratios and periods \citep[using the \texttt{astromet}\footnote{\href{https://github.com/zpenoyre/astromet.py}{https://github.com/zpenoyre/astromet.py}} package developed by][]{Penoyre_2022a}. We assumed no eccentricity and a relation between the mass ($q=M_{\rm{comp}}/M_{\rm{primary}}$) and light ($l=L_{\rm{comp}}/L_{\rm{primary}}$) ratio of the components of $l = q^{3.5}$, a heuristic commonly used for main-sequence stars. We ran $500\,000$ simulations randomly sampling the log-uniform space of mass ratios (from 0.0001 to 1) and periods (from 0.016 to 1585 yr). The mass of the primary source, the viewing angles, and the orbital phase were sampled at random every time following the measured values and uncertainties of stellar mass ($1.30\pm0.08~M_\odot$), inclination ($32\pm1^{\circ}$), and position angle ($10\pm1^{\circ}$, disk geometries from \citealp{Ribas_2023}). We then selected only those simulations which resulted in astrometric properties compatible with what is observed in MP Mus ($0.26>|\Delta \mu|>0.16$ mas~yr$~^{-1}$, $|\Delta \mu|/\sigma_{|\Delta \mu|}>4$). In addition, MP Mus has a re-normalized unit weight error (RUWE) very close to 1 in both \textit{Gaia} DR2 and DR3 (of 0.98 and 0.96, respectively). We hence imposed a conservative threshold to the simulations of $\rm{UWE_{DR3}}<1.1$ \citealp[see section 4.1 of][]{Penoyre_2022a}.

\begin{figure}[h]
    \centering
    \includegraphics[width=\textwidth]{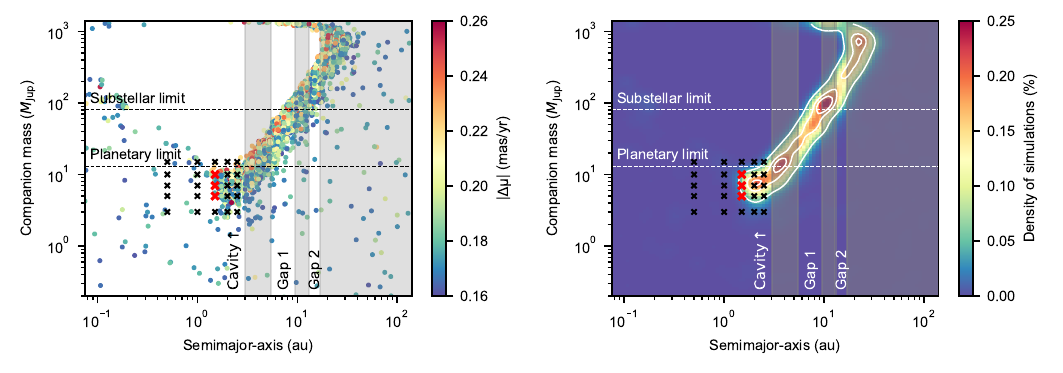}
    \caption{Orbital separation and mass of a companion that could produce the \textit{Gaia} proper motion anomaly of MP Mus. The left panel shows the individual \textit{Gaia} simulations which produce an astrometric signal compatible with the observations ($0.26>|\Delta \mu|>0.16$ mas~yr$~^{-1}$, $|\Delta \mu|/\sigma_{|\Delta \mu|}>4$, and $\rm{UWE_{DR3}}<1.1$). The right panel presents the same simulations but smoothed over a 2d histogram (contours correspond to 0.05, 0.15, and 0.2\,\%). When considering the disk structure traced in the millimetre (shaded area), the most likely companion is a gas giant at 1-2~au. Crosses indicate the parameters used for the hydrodynamical simulations in Sec.~\ref{sec:simulations}. The simulations that best reproduce both \textit{Gaia} and ALMA observations are show in red (Sec. \ref{sec:discussion}).}\label{fig:PMa}
    \end{figure}

Figure~\ref{fig:PMa} shows the orbital separations and masses for a companion to induce the observed PMa and RUWE. We note that we only model a two-body interaction, and the interpretation of the PMa becomes more complex for multi-body systems. In this work we assume that, although there may be other planets in MP~Mus, its PMa is dominated by a single companion. From the \textit{Gaia} data, we conclude that the PMa of MP~Mus is most likely due to a companion at orbital separations $<$30~au, with the required mass increasing with orbital distance (Fig.~\ref{fig:PMa}). The non-detection of astrometric noise (i.e., RUWE$\approx$1) also rules out massive bodies with short periods (i.e., $a<1$~au). The strong statistical significance of the \textit{Gaia} PMa detection makes for a good companion candidate in MP~Mus, but a second, independent signature is needed to further constrain its location and mass.

\subsection{Hidden substructures at 3~mm}\label{sec:alma_3mm}

To better understand MP~Mus we obtained new ALMA observations at 3~mm and with a spatial resolution (4~au) similar to that of the 1.3~mm data presented by \citep{Ribas_2023} (see Methods for details about the observations and data processing). Longer wavelengths offer two main advantages to search for substructures in disks. First, the dust opacity (and therefore the optical depth) decreases with wavelength, and the observations probe emission closer to the disk midplane where planets are expected to form. This is most important in the disk inner regions where the emission is likely to be optically thick, as indicated by the low ($\alpha<2$) local millimetre spectral index of MP~Mus between 1.3 and 2.2~mm \citep{Ribas_2023}. A second benefit of observations at longer wavelengths is that they trace larger grains, which experience a stronger gas drag and are more strongly trapped in pressure maxima. Therefore, if present, substructures may show an enhanced contrast at longer wavelengths \citep[e.g.,][]{Pinilla_2015}.

\begin{figure}[h]
    \centering
    \includegraphics[width=\textwidth]{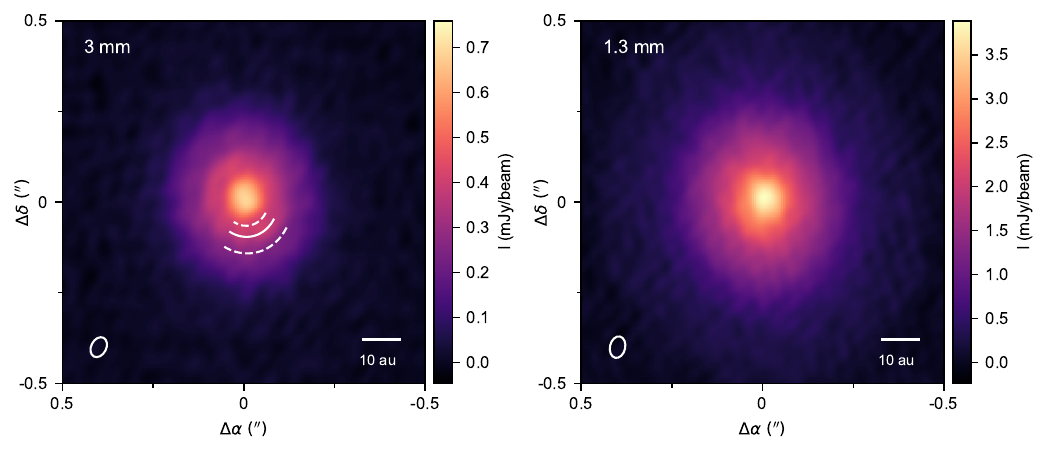}
    \caption{ALMA continuum observations of MP~Mus at 3~mm and 1.3~mm. The solid and dashed white arcs in the 3~mm image mark the ring and two gaps. The beam of each observation is indicated as a white ellipse in the bottom left corners.}\label{fig:ALMA}
    \end{figure}

\begin{figure}[h]
    \centering
    \includegraphics[width=0.9\textwidth]{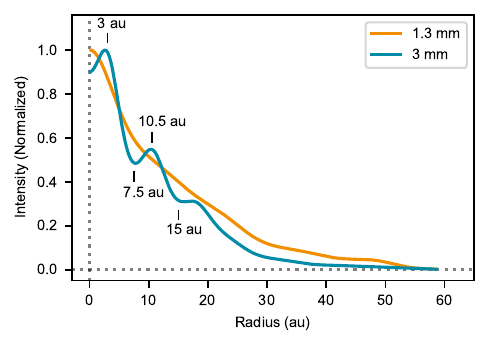}
    \caption{Intensity radial profiles of MP~Mus derived using \textsc{frank}. The 3~mm profile shows structures undetected at shorter wavelengths, including an inner cavity, a ring, and two gaps. These features are indicated in the figure.}\label{fig:rad_profiles}
    \end{figure}

The ALMA 3~mm observations are shown in Fig.~\ref{fig:ALMA} alongside with previous 1.3~mm data. The disk appears smooth at 1.3~mm \citep{Ribas_2023}, but the new 3~mm data reveal a ring at 10.5~au and two gaps at 7.5 and 15~au that are undetected at shorter wavelengths. The 3~mm image also has flatter central emission, with hints of two peaks suggesting the presence of a barely resolved cavity (this is better seen in Supplementary Fig.~1). To better characterise the structure of the system, we calculated and compared the intensity radial profiles of the disk at both wavelengths. We do this in visibility space with the code \textsc{frank} \citep{frank}, which achieves higher angular resolution and recovers smaller structures than working with synthesised images. The profiles are shown in Fig.~\ref{fig:rad_profiles} and more clearly reveal the aforementioned structures in the 3~mm image: an inner cavity with a radius of $\sim$3~au, as well as a ring and two gaps. The \textsc{frank} residuals are provided in Supplementary Fig.~2. Although not as clear, radial profiles calculated directly from synthesised images also show these substructures (see Methods and Supplementary Fig.~3).

The 3~mm data prove that MP~Mus is indeed structured, which could explain both the disk longevity and the similar ratio of gas/dust radii to other structured disks. A naive interpretation of the disk architecture, in which the inner cavity and two gaps are each being carved by a different planet, would imply that MP~Mus hosts a multi-planetary system. Nevertheless, we caution that mechanisms other than planets can open cavities and gaps \citep{Zhang_2015,Flock_2015,Gonzalez_2017} and that such interpretation may be too simplistic. Moreover, simulations have shown that a single planet may carve multiple gaps in low viscosity discs \citep{Dong_2017, Bae_2017}, although this is no longer the case for models including radiative effects \citep{Ziampras_2020}.

We can, however, attempt to link the new-found structures to the \textit{Gaia} PMa to search for possible explanations for both phenomena. First, previous observing campaigns to detect planets and substellar objects in MP~Mus have ruled-out sources more massive than $2-3~M_{\rm Jup}$ at separations $>40-50$~au \citep[e.g.,][]{Wolff_2016,AsensioTorres_2021}. If instead we assume that a putative companion in either of the gaps (i.e., at 7.5~au or at 15~au) is responsible for the PMa, it would require masses above $10~M_{\rm Jup}$ to account for the observed signal (Fig.~\ref{fig:PMa}). Such a massive object would open much wider gaps and strongly alter the disk structure, in stark contrast with its narrow gaps and overall smooth appearance \citep[for comparison, the vast majority of planetary masses determined from gap properties are well below 3~$M_{\rm Jup}$,][]{Bae_2023, Ruzza_2024}. For these reasons, even if planets may be carving those gaps, they are unlikely to be the cause of the PMa. On the other hand, the inner $\sim$3~au cavity appears as an optimal location for possible companions. We can further restrict the location of the companion if we assume that it is responsible for carving the cavity: even in the most favourable scenario (i.e., an equal mass binary), a non-eccentric binary system can only open a cavity as large as 2-3 times the orbital separation \citep{Artymowicz_1994,Ragusa_2020}. Put together, the \textit{Gaia} and ALMA data strongly suggest that MP~Mus harbours a giant protoplanet in the inner 1-3~au of the system.

\subsection{Confronting observations with simulations}\label{sec:simulations}

\begin{figure}[h]
    \centering
    \includegraphics[width=1.0\textwidth]{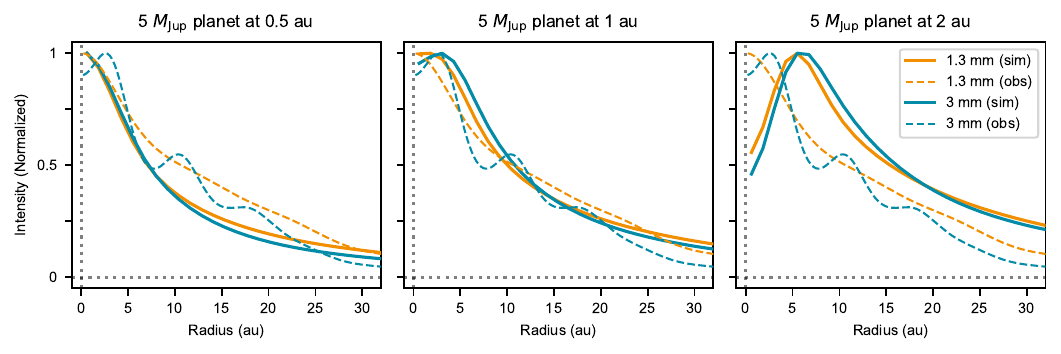}
    \caption{Intensity radial profiles for three of the SPH simulations. The simulation profiles are shown as solid lines at 1.3~mm (orange) and 3~mm (blue), after convolution with the beams of the ALMA observations. The observed profiles are also shown as dashed lines. The panels show a 5~$M_{\rm Jup}$ planet at  0.5~au (left), 1~au (middle), and 2~au (right). In the first case, the cavity opened by the planet is too small to be resolved in the observations. Planets with separations $\geq$2~au produce cavities that are too large to be compatible with the data.}\label{fig:rad_profiles_sim}
    \end{figure}

To explore if a protoplanet within the mass range and orbital separations suggested by the \textit{Gaia} PMa could explain the inner cavity seen at 3~mm, we ran hydrodynamical simulations of a star + planet system using the smoothed particle hydrodynamics (SPH) code \textsc{Phantom} \citep{Price_2018}. We ran a grid of 25 models with different orbital radii and masses that would fit inside the 3~au cavity, namely at 0.5, 1, 1.5, 2, and 2.5~au (corresponding to orbital periods of 0.3, 0.9, 1.6, 2.5, and 3.5 years, respectively), and with companion masses of 3, 5, 7, 10, and 15~$M_{\rm Jup}$. Details on the modelling setup are provided in the Methods section. We then performed radiative transfer with the \textsc{MCFOST} code \citep{MCFOST,MCFOST2} to produce continuum images at 1.3 and 3~mm and convolved the results with the corresponding ALMA beams. Fig.~\ref{fig:rad_profiles_sim} shows the radial profiles of the simulations at {1.3 and 3~mm} for three representative cases as calculated directly from the images with the code \textsc{GoFish} \citep{GoFish}. The results for the full grid are provided in Section 2 of the Supplementary Material and in Supplementary Fig.~4.

The SPH simulations provide additional information about the properties of the companion. First, none of the models with an orbital separation of 0.5~au carve an inner cavity large enough to be discernible with the current spatial resolution, regardless of their mass. On the other extreme, planets at $\geq$2~au produce cavities that are too large to be compatible with the observations. The mass of the companion does not have a strong effect on the cavity size, but more massive objects result in deeper cavities. We also note that some simulations create a cavity of the right size at 3~mm while no cavity is visible at 1.3~mm, just as observed with ALMA (although this is more sensitive to the adopted disk and dust properties). Overall, the simulations show that the observed cavity can be successfully explained by a gas giant orbiting at 1-2~au, lending further indirect but independent support to the presence of such planet in MP~Mus.

\subsection{Discussion}\label{sec:discussion}

The combination of two independent but consistent pieces of evidence (the \textit{Gaia} PMa and the finding of a small inner cavity) create a very compelling case for the existence of MP~Mus~b, a gas giant planet orbiting between $\sim$1-3~au. The case is strengthened even further when we consider the SPH simulations of the system, which show that planets between 1-2~au carve a cavity with the correct size, and in some cases even reproduce the peculiar feature of the inner cavity being visible at 3~mm but not at 1.3~mm. The presence of a companion in the inner regions of MP~Mus would also account for its low accretion rate compared to other disks with similar masses \citep[$\sim10^{-10}$~$M_\odot$ yr$^{-1}$,][]{Ingleby_2023}. It is unlikely that such a planet would form at its current location: planet formation is thought to be more efficient at radii near and outside the snowline (the location at which water ice sublimates) due to the change in dust properties and enhanced dust-to-gas ratios \citep[e.g.,][]{Drkazkowska_2017}, and the gas giants found with 1-2~au orbits are believed to have undergone planetary migration \citep[e.g.][]{Alexander_2012}. Interestingly, at 1-2~au, it is possible that MP~Mus~b resides in the habitable zone of its 1.3~$M_\odot$ star. Detecting such a planet with radial velocity would be a formidable challenge, as the expected signal of a $\sim$5~$M_{\rm Jup}$ between 1 and 3~au (i.e., a radial velocity semi-amplitude of 40-70~m~s$^{-1}$) is below the typical level of stellar activity in young stars \citep[$\sim$100~m~s$^{-1}$, e.g.,][]{Zakhozhay_2022}, and even detections of very massive protoplanets much closer to their host stars are very difficult to disentangle from other sources of radial velocity signal \citep[see the case of CI~Tau,][]{Johns-Krull_2016, Biddle_2018, Donati_2020, Donati_2024, Manick_2024}. This highlights the unique potential of the PMa technique to uncover the population of young planets in close (a few au) orbits embedded in protoplanetary disks, a region of the parameter space that remains beyond the reach of other current detection methods. This becomes especially promising when combined with additional information about the possible presence of the protoplanet such as in the case of MP~Mus, where the disk structure traced by ALMA allows to pinpoint the location of the companion; the narrow width of the two gaps rules them out as possible sites for the body inducing the PMa, leaving the inner cavity as the only suitable location.

In addition to the inner cavity revealed by the new 3~mm data, the different morphology of the disk at 1.3 and 3~mm can be solely attributed to the difference in wavelength (given that both observations have the same spatial resolution). As mentioned in Sec.~\ref{sec:alma_3mm}, this arises from a combination of the enhanced contrast of substructures (due to more efficient trapping of larger grains) and the lower optical depth with increasing wavelength. Previous studies have already shown that disk substructures often appear sharper and with a higher contrast at longer wavelengths \citep[e.g.,][]{CarrascoGonzalez_2019,Huang_2020,Doi_2023}. MP~Mus represents an extreme example of this as it is the first source in which new substructures are uncovered at longer wavelengths, but it is unlikely to be the only system in which this occurs. The new 3~mm observations also confirm that 1~mm emission from the inner regions of protoplanetary disk can remain optically thick and, more importantly, indicate that our current census of disk substructures \citep[see][and references therein]{Bae_2023} is potentially very incomplete. Protoplanetary disks may be more structured than currently believed, on the one hand providing even more dust trapping of large grains, and on the other hand clearly necessitating high-resolution, high-sensitivity surveys at longer wavelengths to understand the frequency and properties of a potential population of hidden substructures.

\section{Methods}\label{Methods}

\subsection{ALMA data processing and imaging}\label{sec:ALMA}

The MP~Mus ALMA Band~3 (3~mm) data correspond to the ALMA project 2022.1.01758.S (P.I.: Á. Ribas) and were taken with two different configurations. The compact configuration was observed once on 21 May 2022 and contains baselines from 27~m to 3.6~km. The extended configuration was observed three times on 27-28 July 2022 with baselines ranging from 230~m to 16.2~km. The correlator was setup to maximise the continuum sensitivity, containing four spectral windows centred at 90.521, 92.416, 102.521, and 104.479~GHz, each with a bandwidth of 1.875~GHz and 128 channels. All the observations were first calibrated using the pipeline script and \textsc{CASA v.6.4.1.12}, and further processing was performed with \textsc{CASA v.6.5.6.22}. In particular, we first performed phase-only self-calibration on the extended and compact configurations separately. We then placed the phase centre on the disk for each observation, set the coordinates of each phase centre to a common value, and re-scaled the flux of the three extended executions to match the flux of the compact one. Finally, one additional round of phase-only self-calibration was applied after combining the four execution blocks.

The Band~6 data (1.3~mm) are part of ALMA project 2017.1.01167.S (P.I.: S. Pérez) and were already presented in \citep{Ribas_2023}. In that work, \citep{Ribas_2023} combined these observations with those of ALMA project 2017.1.01419.S (P.I.: C. Cáceres), which had lower spatial resolution but were useful to study gas emission in the system. We chose not to include those data in this work since here we focus on the continuum only. The data consist of a compact configuration observed on 15 January 2018 with baselines between 15~m and 2.4~km, and an extended one taken on 16 November 2017 with baselines between 90~m and 8.2~km. The observations were calibrated with the pipeline and \textsc{CASA v.5.1.1-5}, and we then used \textsc{CASA v.6.5.6.22} for the rest of the analysis. The correlator setup in this case contained four spectral windows, three for continuum centred at 232.483, 244.983, and 246.983~GHz (bandwidth of 1.875~GHz and 128 channels each), and the fourth one centred at 230.525~GHz targeting the $^{12}$CO (2-1) line (bandwidth of 1.875~GHz and 960 channels). We flagged channels containing CO emission and then performed self-calibration with a similar process as for the Band~3 data: we first self-calibrated the compact and extended configurations individually, placed the phase centre at the centre of the disk in each case and changed its coordinates to a common value, and then re-scaled the compact configuration to have the same flux as the extended one (in this case, the observatory calibrator appeared more stable around the time of the extended observations, although this has no impact on any of our results). We then ran one final round of phase-only self-calibration in which both the compact and extended configurations were combined.

We imaged the data using the {\tt tclean} routine in \textsc{CASA} using different weighting to achieve the best compromise between spatial resolution and sensitivity. For the purpose of analysing the disk structures (Fig.~\ref{fig:ALMA}) we used the {\tt mtmfs} option with {\tt nterms=1} and scales of 0, 1, 3, and 5 beams. We adopted a {\tt robust} parameter of 0 for both wavelengths, resulting in a beam of 0.06"$\times$0.04" in both cases, and a position angle of $PA=-13^\circ$ at 1.3~mm and $-27^\circ$ at 3~mm. The corresponding image rms values are 37~$\mu$Jy~beam$^{-1}$ and 7.7~$\mu$Jy~beam$^{-1}$ at 1.3 and 3~mm, respectively. The peak S/N of these images is 105 at 1.3~mm and 90 at 3~mm. The disk flux, radius, and mass measured at 3~mm are provided in Section 1 of the Supplementary Material.

\subsection{Radial profiles and additional sanity checks}\label{sec:rad_profiles}

The radial profiles in Fig.~\ref{fig:rad_profiles} were created using the \textsc{frank} code \citep{frank}. Under the assumption that the the disk is axisymmetric, this software calculates the radial profile of the disk directly from the observed visibilities using a Hankel transform. For both the 1.3 and 3~mm data we first fit the geometry of the system with the non-parametric option and using the measured disk inclination and position angles as initial guesses \citep[$i$=32$^\circ$, $PA$=10$^\circ$,][]{Ribas_2023}. The values of $i$ and $PA$ determined by \textsc{frank} are compatible with the 1$^\circ$ uncertainty of these quantities. We then explored the effect of changing the $\alpha$ and $w_{\rm smooth}$ hyper-parameters: the results were largely insensitive to the choice of these in the case of the 1.3~mm data, yielding a profile similar to the one presented in \citep{Ribas_2023} (unless extreme values of $\alpha$ and $w_{\rm smooth}$ were used, which induced artificial wiggles) and we adopted $\alpha$=1.2 and $w_{\rm smooth}=5\times10^{-3}$. In the case of 3~\,mm we found that going below $\alpha$=1.3 or $w_{\rm smooth}=10^{-2}$ produced clearly artificial ripples in the profile, and we thus used these values. We then imaged the residuals after subtracting the \textsc{frank} visibilities from the observed ones (Supplementary Fig.~2), which appear largely signal-free: the residuals are below three times the image noise (rms) in the case of the 3~mm observations, and the 1.3~mm show only two localised blobs around the 5~rms level.

We performed two sanity checks to ensure that the observed structures are not an artefact from \textsc{frank}. First, we also calculated the radial profiles directly from the synthesised images using the \textsc{GoFish} code \citep{GoFish}. These profiles have less spatial resolution (image synthesis incurs some information loss during the binning and weighting of visibilities), but serve to confirm that substructures are present in the 3~mm image and are not an artefact of \textsc{frank}. For this purpose, we produced images with {\tt robust=-0.5} to improve the angular resolution. The ring and two gaps are also visible in the image profile. The cavity is only tentatively detected in the profile as a flattening of the intensity in the inner regions. 

In addition, we also used the non-parametric \textsc{gpuvmem} image reconstruction code \citep{Carcamo_2018} to create a maximum-entropy version of the image. The maximum-entropy method implemented in \textsc{gpuvmem} belongs to the family of regularised maximum likelihood methods, which have gained significant traction in recent years \citep[e.g.][]{EHT_2019, Zawadzki_2023, mpol}. The self-calibrated continuum data were super-resolved using \textsc{gpuvmem} with no entropy term in the objective function \citep[see][]{Carcamo_2018}; image positivity alone provided sufficient regularisation \citep[see][]{Perez_2020}. This approach yielded an image with slightly higher angular resolution and sensitivity compared to an equivalent {\tt tclean} image. The \textsc{gpuvmem} image reproduces the same substructures identified in the {\tt tclean} image while also revealing a small central cavity.

These tests confirm that the inner cavity, ring, and gaps recovered by \textsc{frank} are real features. The 3~mm {\tt robust=-0.5} and the \textsc{gpuvmem} images, as well as the corresponding radial profiles, are provided in the Supplementary Figs.~2 and~3.

\subsection{PHANTOM simulations}\label{sec:PHANTOM}

\subsubsection{Hydrodynamical simulations}

We ran 3D global hydrodynamical simulations using the SPH code \textsc{Phantom} \citep{Price_2018}, widely used to study gas and dust dynamics in planet forming environments \citep{Dipierro_2015,Toci_2020,Veronesi_2020}. To model the inner cavity of MP~Mus, we ran a grid of 25 simulations varying the semi-major axis and the mass of the companion within ranges compatible with the \textit{Gaia} PMa. In all the simulations, the mass of the star is $M_\star = 1.3~M_\odot$ and the disk mass is $M_{\rm disk} = 0.01~M_\odot$ (informed by the dust mass calculated from the 3~mm emission and adopting a standard gas-to-dust ratio of 100, see section~1 of the Supplementary Material). The disk is locally isothermal with a mid-plane temperature profile $T\propto R^{-0.5}$, and the initial surface density profile follows $\Sigma \propto R^{-1}$. We adopted a typical disk aspect ratio $H/R$ to 0.1 at a reference radius of 100~au \citep{Law_2021, Paneque_2023}. The resolution is set by the number of particles ($N=10^6$). The outer radius is initialised to 50~au as traced by the 1.3~mm continuum observations, and the inner radius is chosen according to the semi-major axis of the inner companion $a_{\rm p}$, namely $R_\text{in} = 1.5~a_{\rm p}$. The central star and the companion were modelled as sink particles, with accretion radii $R_{\rm acc, \star}=0.75~a_{\rm p}$ for the central star and $R_{\rm acc, p}=0.25R_h$ for the secondary, where $R_{\rm h}$ is the companion Hill radius. The choice of the primary sink radius is safe since it is always smaller than the semi-major axis of the companion, and the radius of cavity we observe is always 3-4 times bigger, meaning that it is carved by the tidal interaction rather than being a numerical artefact. We used SPH shock capturing viscosity to model the mechanisms transporting the angular momentum through the disk, prescribing shock capturing viscosity coefficients $\alpha_\text{AV} = 0.2$ and $\beta_\text{AV} = 2$ which resulted in a viscosity parameter of $\alpha_\text{\rm ss}= 5\times10^{-3}$. 

The dust component is set by adopting an initial gas-to-dust ratio of 100 uniformly across the entire disk. We simulated five dust grains sizes ranging from $1~\mu$m to 0.5~cm and with intrinsic density 3~g~cm$^{-3}$. The dust is treated with the mixture algorithm, also known as one fluid algorithm (Laibe Price), suitable for coupled dust grains. The mass of the five different grain size populations is calculated according to $n(s) \propto s^{-3.5}$, where $s$ is the grain size.

The mass and semi-major axis of the companion are $M_p\in(3,5,7,10,15)~ M_{\rm Jup}$ and $a_p\in(0.5,1,1.5,2,2.5)$~au, as discussed in section \ref{sec:PMa}. We let the simulations evolve for $1000$ planet orbits, enough time to ensure that the inner cavity reaches a steady state \citep{Ragusa_2020}.

\subsubsection{Radiative transfer}

The thermal structure and the 1.3~mm and 3~mm continuum emission images were computed using the Monte Carlo radiative transfer code \textsc{MCFOST} \citep{MCFOST,MCFOST2}. The code takes as input the distribution of gas and dust grains from the hydrodynamical simulation to create a Voronoi mesh where each cell corresponds to an SPH particle. The disk receives passive heating from the central star, modelled as a black body with temperature of 5000~K, radius $R_\star$=1.25~$R_\odot$ and mass $M_\star$=1.3~$M_\odot$. We adopted a DIANA standard dust composition \citep{Woitke_2016, Min_2016} – i.e. 60\% silicates, 15\% amorphous carbon \citep[optical constants from][respectively]{Dorschner_1995, Zubko_1996} and 25\% porosity. The code assumes that sub-$\mu$m grains follow the gas, while the spatial distribution of dust grains between $\mu$m and cm is obtained by interpolation. We discretise the radiation field using $N_\gamma = 10^8$ photons in order to compute the temperature structure of the disk assuming local thermal equilibrium. Images are then computed via ray-tracing again with $N_\gamma = 10^8$ photons.

\section{Data availability}

All the data used in this work are publicly available through the ALMA archive (programs 2022.1.01758.S and 2017.1.01167.S) and the \textit{Gaia} archive.

\section{Code availability}

The Python packages used in this study are publicly available:\\
\textsc{astromet} (\href{https://github.com/zpenoyre/astromet.py}{https://github.com/zpenoyre/astromet.py}),\\
\textsc{frank} (\href{https://github.com/discsim/frank}{https://github.com/discsim/frank}),\\
\textsc{PHANTOM} (\href{https://github.com/danieljprice/phantom}{https://github.com/danieljprice/phantom}),\\
\textsc{MCFOST} (\href{https://github.com/cpinte/mcfost}{https://github.com/cpinte/mcfost}),\\
and \textsc{gpuvmem} (\href{https://github.com/miguelcarcamov/gpuvmem}{https://github.com/miguelcarcamov/gpuvmem}).

\section{Acknowledgements}

We thank the reviewers for their careful analysis of this work, which has significantly improved the manuscript. We also thank Marion Villenave and Stefano Facchini for insightful discussions and comments, and Miguel Cárcamo for his help with running {\tt gpuvmem}. A.R., C.L, and C.J.C. have been supported by the UK Science and Technology Facilities Council (STFC) via the consolidated grant ST/W000997/1. This project has received funding by the European Union’s Horizon 2020 research and innovation programme under the Marie Sklodowska-Curie grant agreements No. 823823 (RISE DUSTBUSTERS project). N.C. acknowledges funding from the European Research Council (ERC) under the European Union Horizon Europe programme (grant agreement No. 101042275, project Stellar-MADE). S.P. acknowledges support from FONDECYT grant 1231663 and funding from ANID -- Millennium Science Initiative Program -- Center Code NCN2024\_001. I.dG. acknowledges support from grant PID2020-114461GB-I00, funded by MCIN/AEI/10.13039/501100011033. This paper makes use of the following ALMA data: ADS/JAO.ALMA\#2022.1.01758.S and ADS/JAO.ALMA\#2017.1.01167.S. ALMA is a partnership of ESO (representing its member states), NSF (USA) and NINS (Japan), together with NRC (Canada), MOST and ASIAA (Taiwan), and KASI (Republic of Korea), in cooperation with the Republic of Chile. The Joint ALMA Observatory is operated by ESO, AUI/NRAO and NAOJ.

\section{Author contributions}

A.R. led the ALMA Band 3 proposal, processed and analysed the ALMA observations, led the project, and wrote the manuscript. M.V. performed the analysis of the \textit{Gaia} proper motion anomaly of MP~Mus. F.Z. and E.M. aided with the processing of the ALMA data. C.L. ran the PHANTOM simulations. S.P. produced the {\tt gpuvmem} images of MP~Mus. All authors contributed to the interpretation of the results and commented on the manuscript.

\section{Competing interests}

The authors declare no competing interests.

\section{Supplementary material}

\subsection{Properties and additional analysis of the ALMA data at 3~mm}

We measured a 3~mm flux of 22$\pm$1~mJy from the image synthesised with {\tt robust=0.5}. The uncertainty is largely dominated by the absolute calibration (5\% in Band~3, see the ALMA Technical Handbook). We then calculated a dust disk mass of $0.12^{+0.03}_{-0.02}$~$M_{\rm Jup}$ using standard assumptions \citep[e.g.,][]{Andrews_2005}: optically thin emission, an average dust temperature of 20~K, and a power-law dependence of the dust opacity with frequency following $\kappa_\nu = \kappa_\text{230 GHz} (\nu/\text{230 GHz})^\beta$, where $\kappa_\text{230 GHz}$=2.3 cm$^2$~g$^{-1}$ and $\beta$ values between 0 and 0.5, as typically inferred in protoplanetary disks \citep{Tazzari_2021}. To determine the disk mass and its uncertainty we bootstrapped 1000 estimates by varying the 3~mm flux within its uncertainty, as well as the temperature value following a Normal distribution with a standard deviation of 2.5~K, and $\beta$ with a uniform distribution between 0 and 0.5. The reported dust mass and uncertainties correspond to the median and the 16\% and 84\% values. This mass is compatible with the dust mass derived from the 1.3~mm flux in \citep{Ribas_2023}.

The radii encompassing 68\%, 90\%, and 95\% of the 3~mm continuum flux are 22$\pm$5~au, 35$\pm$5~au, and 44$\pm$5~au respectively, based on the cumulative curve of the intensity profiles. We used the {\tt robust=0} images for this purpose as they provide a good compromise between angular resolution and sensitivity, but the results with other weightings are compatible within beam size of the observations. The comparison with the $R_{68\,\%}$ and $R_{90\,\%}$ values at 1.3~mm reported in \citep{Ribas_2023} also shows that the disk appears $\sim$30\% more compact at 3~mm.

\begin{figure}[h]
    \centering
    \includegraphics[width=0.48\textwidth]{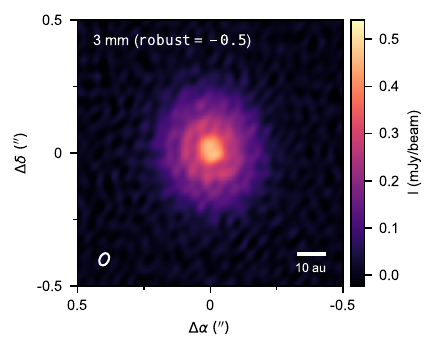} \includegraphics[width=0.48\textwidth]{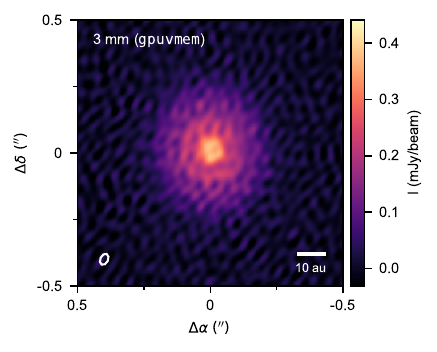}
    \caption{ALMA 3~mm observations of MP~Mus synthesised with higher angular resolution. The left panel shows the result of the \textsc{tclean} algorithm with {\tt robust=-0.5}, and the image using the \textsc{gpuvmem} code after convolution with a similar beam is shown on the right panel. This results in noisier images but improves the angular resolution (0.05"$\times$0.03"). The inner cavity, ring, and gaps are more clearly visible.}\label{fig:B3_robust-0.5}
    \end{figure}

\begin{figure}[h]
    \centering
    \includegraphics[width=\textwidth]{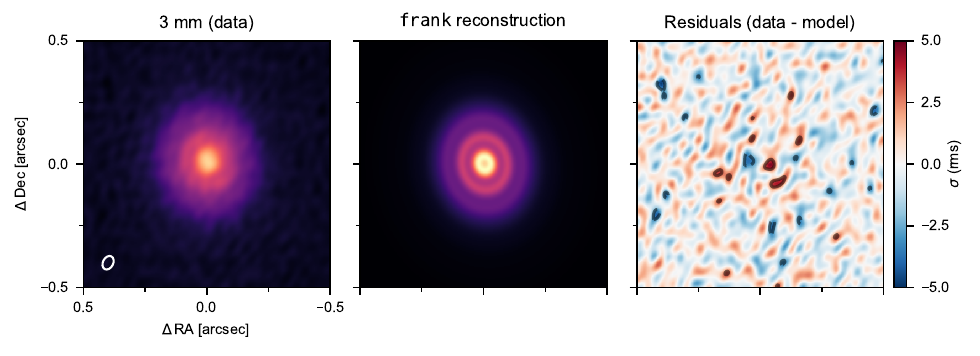}
    \includegraphics[width=\textwidth]{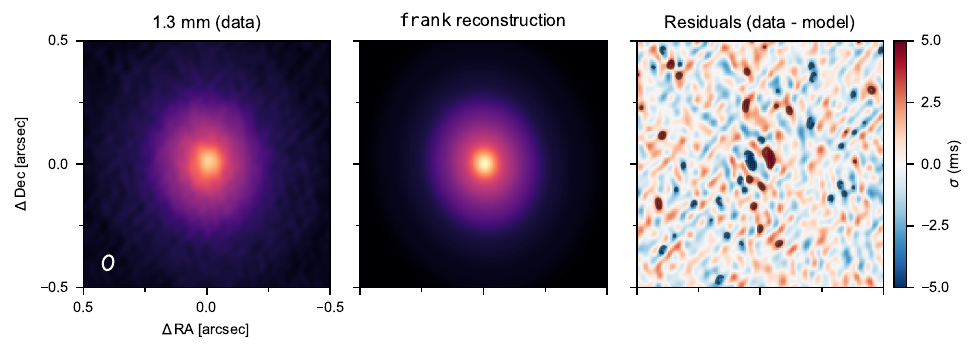}
    \caption{Residuals of the \textsc{frank} radial profiles at 3~mm and 1.3~mm. For each case (3~mm top row, 1.3~mm bottom row) the observations are shown on the left, the axisymmetric disk reconstructed from the \textsc{frank} profile in the middle, and the residuals appear on the right. Contours are show at $\pm$3 and $\pm$5 times the noise level. Only two localised blobs of $\sim$5~rms appear in the 1.3~mm data, while no residuals are found above 3~rms in the 3~mm case.
    }\label{fig:frank_residuals}
    \end{figure}

\begin{figure}[h]
    \centering
    \includegraphics[width=0.9\textwidth]{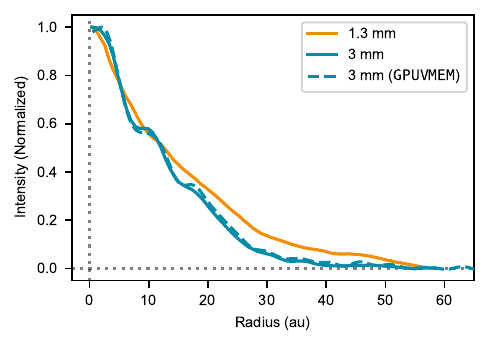}
    \caption{Intensity radial profiles of MP~Mus at 1.3 and 3~mm calculated directly from images. The profiles are calculated using the {\tt robust=-0.5} image (solid lines) and the image from \textsc{gpuvmem} (dashed line) are shown. While the profile of the 1.3~mm appears mostly smooth, the 10.5~au ring and the two gaps at 7.5 and 15~au are clearly visible in the 3~mm one. Although not as clear, the presence of the inner cavity at 3~mm is reflected by the flattening of the radial profile in the inner region.}\label{fig:rad_profiles_image}
    \end{figure}

The 3~mm {\tt robust=-0.5} image using {\tt tclean} in {\tt CASA} and the one produced by \textsc{gpuvmem} and convolved with a similar beam are presented in Fig.~\ref{fig:B3_robust-0.5}. Fig.~\ref{fig:frank_residuals} shows the residuals of the \textsc{frank} fit for both wavelengths. These were obtained by first subtracting the \textsc{frank} visibilities from the observed ones, and then imaging the result with {\tt robust=0.0}. Finally, the radial profiles calculated from the {\tt robust=-0.5} and \textsc{gpuvmem} images using \textsc{GoFish} \citep{GoFish} are shown in Fig.~\ref{fig:rad_profiles_image}.

\subsection{Full set of PHANTOM simulations}

\begin{figure}[h]
    \centering
    \includegraphics[width=\textwidth]{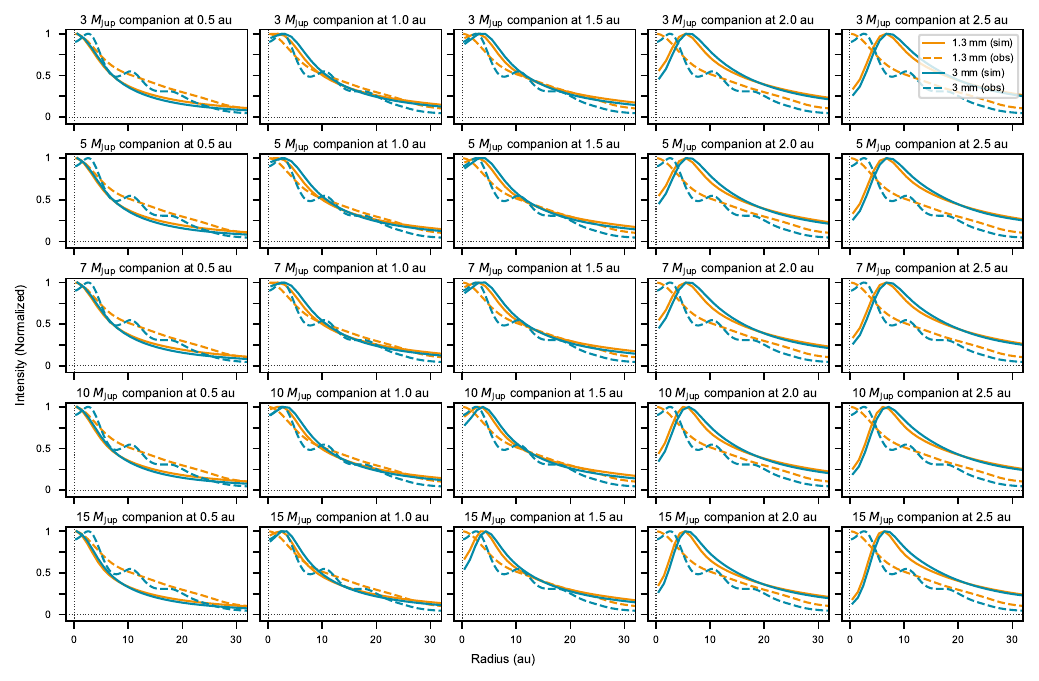}

    \caption{Intensity radial profiles for the full grid of PHANTOM models. Profiles for the simulations (solid lines) and the observations (dashed lines) are shown at 1.3~mm (orange) and 3~mm (blue).}\label{fig:PHANTOM_full}
    \end{figure}

Figure~\ref{fig:PHANTOM_full} shows the corresponding radial profiles for the full grid of PHANTOM models. As mentioned in Section~\ref{sec:PHANTOM}, companions at separations $<$1~au do not open a large enough cavity, while separations $\geq$2~au carve cavities that are too wide. Increasing the mass of the planet results in deeper cavities. It is also worth noticing that a 3-7~$M_{\rm Jup}$ companion at $\sim$1~au creates a cavity at 3~mm that remains undetected at 1.3~mm, just as observed in the ALMA data. However, this is highly sensitive to the disk and dust properties which we have not explored in detail, and should therefore be considered as a strong indication that planets with such properties can explain the observations, rather than as precise measurements of the planet location and mass.

\newpage
\bibliography{bibliography}


\end{document}